\documentclass [12 pt]{article}
\def\be{\begin{equation}}
\def\eea{\end{eqnarray}}
\def\bea{\begin{eqnarray}}
\def\ee{\end{equation}}

\usepackage[psamsfonts]{amssymb}
\usepackage{amsmath}
\author{F. Kheirandish$^{1}$ \footnote{fardin$_{-}$kh@phys.ui.ac.ir} and M.
Amooshahi$^{1}$ \footnote{amooshahi@sci.ui.ac.ir}
\\ $^{1}$ {\small Department of Physics, University of Isfahan,}
\\ {\small Hezar Jarib Ave., Isfahan, Iran.}}
\title{A minimal coupling method for investigating one dimensional
 dissipative quantum systems }
\begin{document}
\maketitle
\begin{abstract}
\noindent Quantum mechanics of a general one dimensional
dissipative system investigated by it's coupling to a
Klein-Gordon field as the environment using a minimal coupling
method. Heisenberg equation for such a dissipative system
containing a dissipative term proportional to velocity obtained.
As an example, quantum dynamics of a damped harmonic oscillator as
the prototype of some important one dimensional dissipative models
investigated consistently. Some transition probabilities
indicating the way energy flows between the subsystems obtained.
\end{abstract}
\section{Introduction}
In classical mechanics dissipation can be taken into account by
introducing a velocity dependent damping term into the equation
of motion. Such an approach is no longer possible in quantum
mechanics where a time-independent Hamiltonian implies energy
conservation and accordingly we can not find a unitary time
evolution operator for both states and observable quantities consistently.\\
To investigate the quantum mechanical description of dissipating
systems, there
 are some treatments, one can consider the interaction
between two systems via an irreversible energy flow [1,2], or take
a phenomenological treatment for a time dependent Hamiltonian
which describes damped oscillations, here we can refer the
interested reader to Caldirola-Kanai Hamiltonian for a damped
harmonic oscillator [3].
\begin{equation}\label{dm1}
H(t)=e^{-2\beta t}\frac{p^2}{2m}+e^{2\beta t}\frac{1}{2}m\omega^2
q^2.
\end{equation}
There are significant difficulties about the quantum mechanical
solutions of the Caldirola-Kanai Hamiltonian, for example
quantizing such a Hamiltonian violates the uncertainty relations
or canonical commutation rules and the uncertainty relations
vanish
as time tends to infinity.[4,5,6,7,8] \\
In 1931, Bateman [9] presented the mirror-image Hamiltonian which
consists of two different oscillator, where one of them
represents the main one-dimensional damped harmonic oscillator.
Energy dissipated by the main oscillator completely will be
absorbed by the other oscillator and thus the energy of the total
system is conserved. Bateman Hamiltonian is given by
\begin{equation}\label{dm2}
H=\frac{p\bar{p}}{m}+\frac{\beta}{2m}(\bar{x}\bar{p}-xp)+(k-\frac{\beta^2}{4m})x\bar{x},
\end{equation}
with the corresponding Lagrangian
\begin{equation}\label{dm3}
L=m\dot{x}\dot{\bar{x}}+\frac{\beta}{2}(x\dot{\bar{x}}-\dot{x}\bar{x})-kx\bar{x},
\end{equation}
canonical momenta for this dual system can be obtained from this
Lagrangian as
\begin{equation}\label{dm3}
p=\frac{\partial L}{\partial
\dot{x}}=m\dot{\bar{x}}-\frac{\beta}{2}\bar{x},\hspace{1.50
cm\bar{p}}=\frac{\partial L}{\partial
\dot{\bar{x}}}=m\dot{x}+\frac{\beta}{2}x,
\end{equation}
dynamical variables $ x, p $ and $\bar{p},\bar{x} $ shoud satisfy
the commutation relations
\begin{equation}\label{dm4}
[x,p]=i,\hspace{2.00 cm} [\bar{x},\bar{p}]=i,
\end{equation}
however the time-dependent uncertainty products obtained in this
way, vanishes as time tends to infinity.[10]\\
Caldirola [3,11] developed a generalized quantum theory of a
linear dissipative system in 1941 : equation of motion of a single
particle subjected to a generalized non conservative force $ Q $
can be written as
\begin{equation}\label{dm5}
\frac{d}{d t}(\frac{\partial T}{\partial \dot{q}})-\frac{\partial
T }{\partial q}=-\frac{\partial V}{\partial q}+Q(q),
\end{equation}
where $ Q_r =-\beta(t)\sum a_{r j}\dot{q}_j $, and $ a_{rj} $'s
are some constants, changing the variable $t$ to $t^{*}$, using
the following nonlinear transformation
\begin{equation}\label{dm6}
t^*=\chi(t),\hspace{1.50cm}dt=\phi(t)dt^*,\hspace{1.50cm}\phi(t)=e^{\int_0^t\beta(t')dt'},
\end{equation}
 together with the definitions
\begin{equation}\label{dm7}
  \dot{q}^*=\frac{d q}{d t^*},\hspace{1.00cm}
L^*=L(q,\dot{q}^*,t^*),\hspace{1.00cm}p^*=\frac{\partial
L^*}{\partial \dot{q}^*},
\end{equation}
the Lagragian equations, can be obtained from
\begin{equation}\label{dm8}
\frac{d}{d t^*}(\frac{\partial L^*}{\partial
\dot{q}^*})-\frac{\partial L^*}{\partial q}=0.
\end{equation}
where $ H^*=\sum p^*\dot{q}^*-L^* $. Canonical commutation rule
and Schrodinger equation in this formalism are
\begin{equation}\label{dm9}
[q,p^*]=i,\hspace{2.00cm}H^*\psi=i\frac{\partial \psi}{\partial
t^*},
\end{equation}
but unfortunately uncertainty relations vanish as time goes to
infinity.[10]\\
Perhaps one of the effective approaches in quantum mechanics of
dissipative systems is the idea of considering an environment
coupled to the main system and doing calculations for the total
system but at last for obtaining observables related to the main
system, the environment degrees of freedom must be eliminated.
The interested reader is referred to the Caldeira-Legget model
[12,13]. In this model the dissipative system is coupled with an
environment made by a collection of $ N $ harmonic oscillators
with masses $ m_n $ and frequencies $\omega_n $, the interaction
term in Hamiltonian is as follows
\begin{equation}\label{dm10}
H'=-q\sum_{n=1}^N c_n
x_n+q^2\sum_{n=1}^N\frac{c_n^2}{2m_n\omega_n^2},
\end{equation}
where  $ q $ and $ x_n $ denote coordinates of system and
environment respectively and the constants $ c_n $ are
called coupling constants.\\
The above coupling is not suitable for dissipative systems
containing a dissipative term proportional to velocity. In fact
with above coupling we can not obtain a Heisenberg equation like $
\ddot{q}+\omega^2 q+\beta\dot{q}=\xi( t)$, for a damped harmonic
oscillator, consistently. In this paper we generalize the
Caldeira-Legget model to an environment with continuous degrees
of freedom by a coupling similar to the coupling between a charged
particle and the  electromagnetic field known as the minimal
coupling. In sections 2, the idea of a minimal coupling method is
introduced .In section 3 the quantum dynamics of a damped harmonic
oscillator is investigated completly and some transition
probabilities indicating the way energy flows between subsystems,
are obtained.In section 4, quantum dynamics of the oscillator
environment is investigated.
\section{Quantum dynamics of a one dimensional damped system}
Quantum mechanics of a one dimensional damped  system  can be
investigated by introducing a reservoir or an environment that
interacts with the system through a new kind of minimal coupling
term. For this purpose let the damped system be a particle with
mass $ m $ influenced by a external potential $ v(q) $. we take
the total  Hamiltonian, i.e., system plus environment like this
\begin{equation} \label{d1}
H=\frac{(p-R)^2}{2m}+v(q)+H_B,
\end{equation}
where  $ q $ and $ p $ are position and canonical conjugate
momentum operators of the particle respectively and satisfy the
canonical commutation rule
\begin{equation}\label{d2}
[q,p]=i,
\end{equation}
and $ H_B $ is the reservoir Hamiltonian
\begin{equation}\label{d3}
H_B(t)=\int_{-\infty}^{+\infty}d^3k  \omega_{\vec{k}}
b_{\vec{k}}^\dag(t) b_{\vec{k}}(t), \hspace{1.50 cm}
\omega_{\vec{k}}=|\vec{k}|.
\end{equation}
Annihilation and creation operators $ b_{\vec{k}}$,
$b_{\vec{k}}^\dag $, in any instant of time, satisfy the following
commutation relations
\begin{equation}\label{d4}
[b_{\vec{k}}(t),b_{\vec{k}'}^\dag(t)]=\delta(\vec{k}-\vec{k}'),
\end{equation}
 and we will show later that reservoir is
a Klein-Gordon type equation with a source term. Operator $ R $
have the basic role in interaction between the system and
reservoir and is defined by
\begin{equation}\label{d5}
R(t)=\int_{-\infty}^{+\infty}d^3k [f(\omega_{\vec{k}})
b_{\vec{k}}(t)+f^*(\omega_{\vec{k}})b_{\vec{k}}^\dag(t)],
\end{equation}
let us call the function $ f(\omega_{\vec{k}}) $, the coupling
function. It can be shown easily that Heisenberg equation for $ q
$ and $ p $ leads to
\begin{eqnarray}\label{d6}
&&\dot{q}=i[H,q]=\frac{\vec{p}-\vec{R}}{m},\nonumber\\
&& \dot{p}=i[H,p]=-\frac{\partial v}{\partial q},
\end{eqnarray}
where after omitting $ p $, gives the following equation  for the
damped quantum system
\begin{equation}\label{d6.5}
m\ddot{q}=-\frac{\partial v}{\partial q}-\dot{R}.
\end{equation}
 Using (\ref{d4}) the Heisenberg equation for $ b_{\vec{k}}$, is
\begin{equation}\label{d7}
\dot{b}_{\vec{k}}=i[H,b_{\vec{k}}]=-i\omega_{\vec{k}}
b_{\vec{k}}+i\dot{q}f^*(\omega_{\vec{k}}),
\end{equation}
with the following formal solution
\begin{equation}\label{d8}
b_{\vec{k}}(t)=b_{\vec{k}}(0)e^{-i\omega_{\vec{k}}
t}+if^*(\omega_{\vec{k}}) \int_0^t d t'
e^{-i\omega_{\vec{k}}(t-t')} \dot{q}(t'),
\end{equation}
substituting $ b_{\vec{k}}(t) $ from (\ref{d8})into (\ref{d6.5})
one can obtain
\begin{eqnarray}\label{d9}
&&m\ddot{q}+\int_0^t d
t'\dot{q}(t')\gamma(t-t')=-\frac{\partial v}{\partial q}+\xi(t)\nonumber\\
&&\gamma(t)=8\pi \int_0^\infty d\omega_{\vec{k}}
|f(\omega_{\vec{k}})|^2\omega_{\vec{k}}^3\cos\omega_{\vec{k}} t\nonumber\\
&&\xi(t)=i\int_{-\infty}^{+\infty} d^3 k
\omega_{\vec{k}}(f(\omega_{\vec{k}})b_{\vec{k}}(0)
e^{-i\omega_{\vec{k}}t}-f^*(\omega_{\vec{k}})b_{\vec{k}}^\dag(0)e^{i\omega_{\vec{k}}t}).
\end{eqnarray}
It is clear that the expectation value of $ \xi(t) $ in any
eigenstate of $ H_B $, is zero. For the following special choice
of coupling function
\begin{equation}\label{d10}
f(\omega_{\vec{k}})=\sqrt{\frac{\beta}{4\pi^2\omega_{\vec{k}}^3}},
\end{equation}
equation (\ref{d9}) takes the form
\begin{eqnarray}\label{d11}
&&m\ddot{q}+\beta\dot{q}=-\frac{\partial v}{\partial q}+\tilde{\xi}(t)\nonumber\\
&&\tilde{\xi}(t)=i\sqrt{\frac{\beta}{4\pi^2}}\int_{-\infty}^{+\infty}
\frac{d^3 k}{\sqrt{\omega_{\vec{k}}}} (
b_{\vec{k}}(0)e^{-i\omega_{\vec{k}}t}-
b_{\vec{k}}^\dag(0)e^{i\omega_{\vec{k}}t}),
\end{eqnarray}
In the following we investigate for example one dimensional
harmonic oscillator.
\section{quantum mechanics of one dimensional damped harmonic
oscillator}
\subsection{quantum dynamics}
For a one dimensional harmonic oscillator with mass $ m $ and
frequency $ \omega $ we have $
 v(q)=\frac{1}{2}m\omega^2q^2 $ and therefore we can
 write (\ref{d11}) as
 \begin{equation}\label{d11.01}
 \ddot{q}+\frac{\beta}{m}\dot{q}+\omega^2q=\frac{\tilde{\xi}(t)}{m}
\end{equation}
with the following solution
\begin{eqnarray}\label{d12}
&&q(t)=e^{-\frac{\beta t}{2m}}(\hat{A}e^{i\omega_1 t}+\hat{B}
e^{-i\omega_1 t})+M(t),\nonumber\\
&& M(t)=i\int_{-\infty}^{+\infty}d^3 k
\sqrt{\frac{\beta}{4\pi^2m^2\omega_{\vec{k
}}}}[\frac{b_{\vec{k}}(0)}{\omega^2-\omega_{\vec{k}}^2-i\frac{\beta}{m}
\omega_{\vec{k}}}e^{-i\omega_{\vec{k}}
t}-\frac{b_{\vec{k}}^\dag(0)}{\omega^2-\omega_{\vec{k}}^2+\frac{i\beta}{m}
\omega_{\vec{k}}}e^{i\omega_{\vec{k}}t}],\nonumber\\
&&
\end{eqnarray}
where $ \omega_1=\sqrt{\omega^2-\frac{\beta^2}{4m^2}}$. Operators
$ \hat{A} $ and $ \hat{B} $, are specified by initial conditions
\begin{eqnarray}\label{d13}
\hat{A}+\hat{B}&=&q(0)-{M}(0),\nonumber\\
(\frac{-\beta}{2m}+i\omega_1)\hat{A}+
(\frac{-\beta}{2m}-i\omega_1)\hat{B}&=&
\dot{q}(0)-\dot{M}(0)\nonumber\\
&=&\frac{p(0)-R(0)}{m}-\dot{M}(0),
\end{eqnarray}
solving above equations and substituting $ \hat{A}$ and $ \hat{B}$
in (\ref{d12}) one obtains
\begin{eqnarray}\label{d14}
&&q(t)=e^{-\frac{\beta t}{2m}}
\{\frac{p(0)}{m\omega_1}\sin\omega_1 t +q(0)\cos\omega_1 t+
\frac{\beta}{2m\omega_1}q(0)\sin\omega_1 t\nonumber\\
&&-\frac{R(0)}{m\omega_1}\sin\omega_1 t -\frac{\beta
M(0)}{2m\omega_1}\sin\omega_1t-M(0)\cos\omega_1 t
-\frac{\dot{M}(0)}{\omega_1}\sin\omega_1 t\}+M(t),\nonumber\\
&&
\end{eqnarray}
also substituting $ q(t) $ from (\ref{d14}) in (\ref{d8}) we can
obtain a stable solution for $ b_{\vec{k}}(t) $ in $
t\rightarrow\infty $ as
\begin{eqnarray}\label{d15}
&& b_{\vec{k}}(t)=b_{\vec{k}}(0)e^{-i\omega_{\vec{k}}
t}-i\sqrt{\frac{\beta }{4\pi^2
\omega_{\vec{k}}^3}}\frac{e^{-i\omega_{\vec{k}}
t}}{(\omega^2-\omega_{\vec{k}}^2-\frac{i\beta}{m}\omega_{\vec{k}})}\{\omega^2
q(0)+i\omega_{\vec{k}}\frac{p(0)-R(0)}{m}\nonumber\\
&&-M(0)\omega^2-i\omega_{\vec{k}}\dot{M}(0)\}\nonumber\\
&&+\frac{i\beta}{4\pi^2m\sqrt{\omega_{\vec{k}}^3}}i
\int_{-\infty}^{+\infty} d^3
\sqrt{\omega_{\vec{k'}}}\{\frac{b_{\vec{k'}}(0)}{\omega^2-\omega_{\vec{k'}}^2-\frac{i\beta}{m}\omega_{\vec{k'}}}
\frac{\sin\frac{(\omega_{\vec{k}}-\omega_{\vec{k'}}t)
}{2}t}{\frac{(\omega_{\vec{k}}-\omega_{\vec{k'}})}{2}}e^{\frac{-i(\omega_{\vec{k}}+\omega_{\vec{k'}})t}{2}}\nonumber\\
&&+\frac{b_{\vec{k'}}^\dag(0)}{\omega^2-\omega_{\vec{k'}}^2+\frac{i\beta}{m}\omega_{\vec{k'}}}
\frac{\sin\frac{(\omega_{\vec{k}}+
\omega_{\vec{k'}})}{2}t}{\frac{(\omega_{\vec{k}}+\omega_{\vec{k'}})}{2}}
e^{\frac{i(\omega_{\vec{k'}}-\omega_{\vec{k}})t}{2}}\},
\end{eqnarray}
now substituting $ b_{\vec{k}}(t) $ from (\ref{d15}) in (\ref{d5})
and using (\ref{d6}), one obtains $ p=m\dot{q}+R $.\\
A vector in fock space of reservoir is a linear combination of
basis vectors
\begin{equation}\label{d15.121}
|N(\vec{k}_1),N(\vec{k}_2),...\rangle_B=\frac{(b_{\vec{k}_1})^{N(\vec{k}_1)}(b_{\vec{k}_2})^{N(\vec{k}_2)}...}{\sqrt{N(\vec{k}_1)!N(\vec{k}_2)!...}}|0\rangle_B
\end{equation}
 where are eigenstates of $ H_B $ and the operators $ b_{\vec{k}} $ and $ b_{\vec{k}}^\dag $ act on them as
\begin{eqnarray}\label{d15.131}
&&b_{\vec{k}}|N(\vec{k}_1),N(\vec{k}_2),...N(\vec{k}),...\rangle_B=\sqrt{N(\vec{k})}|N(\vec{k}_1),N(\vec{k}_2),...N(\vec{k})-1,...\rangle_B\nonumber\\
&&b_{\vec{k}}^\dag|N(\vec{k}_1),N(\vec{k}_2),...N(\vec{k}),...\rangle_B=\sqrt{N(\vec{k})+1}|N(\vec{k}_1),N(\vec{k}_2),...N(\vec{k})+1,...\rangle_B\nonumber\\
&&
\end{eqnarray}
 If the state of system in $ t=0 $ is taken to be $
|\psi(0)\rangle=|0\rangle_B\otimes|n\rangle_\omega $ where $
|0\rangle_B $ is vacuum state of reservoir and $ |n\rangle_\omega
$ an excited state of the Hamiltonian $
H_s=\frac{p^2}{2m}+\frac{1}{2}m\omega^2q^2 $, then it is clear
that
\begin{equation}\label{d16}
\langle\psi(0)| \frac{p^2(0)}{2m}+\frac{1}{2}m\omega^2 q^2(0)
|\psi(0)\rangle=(n+\frac{1}{2})\omega,
\end{equation}
 On the other hand from (\ref{d14}), (\ref{d15}) and (\ref{d5})
we find
\begin{eqnarray}\label{d16.5}
&& lim_{t\rightarrow\infty}[\langle\psi(0)| :
\frac{1}{2}m\dot{q}^2+\frac{1}{2}m\omega^2 q^2
: |\psi(0)\rangle]= 0,\nonumber\\
&&lim_{t\rightarrow\infty}[ \langle\psi(0)| :
\frac{p^2}{2m}+\frac{1}{2}m\omega^2 q^2
:|\psi(0)\rangle]\nonumber\\
&&=\frac{\beta^2\omega^4}{2\pi^2
m}lim_{t\rightarrow\infty}|\int_{-\infty}^{+\infty}\frac{d
x}{x}\frac{e^{ixt}}{\omega^2-x^2+i\frac{\beta}{m}x}|^2\langle q^2(0)\rangle_n\nonumber\\
&&\simeq\frac{\beta^2}{2m}\langle
q^2(0)\rangle_n=\frac{\beta^2}{2m^2\omega}(n+\frac{1}{2})
\end{eqnarray}
where  $:\hspace{01.00 cm}  : $ denotes normal ordering operator.
Now by substituting $ b_{\vec{k}}(t) $ from (\ref{d15}) into
(\ref{d3}), it is easy to show that
\begin{eqnarray}\label{d17}
&&lim_{t\rightarrow\infty}[ \langle\psi(0)|
: H_B(t): |\psi(0)\rangle]\nonumber\\
&&=\frac{\beta\omega^4}{\pi}\int_0^\infty \frac{d
x}{(\omega^2-x^2)^2+\frac{\beta^2}{m^2}x^2}\langle
q^2(0)\rangle_n+\frac{\beta}{\pi m^2 }\int_0^\infty \frac{x^2d x
}{(\omega^2-x^2)^2+\frac{\beta^2}{m^2}x^2}\langle
p^2(0)\rangle_n\nonumber\\
&&=\frac{\beta\omega^3}{\pi m}(n+\frac{1}{2})\int_0^\infty \frac{d
x}{(\omega^2-x^2)^2+\frac{\beta^2}{m^2}x^2}+\frac{\beta\omega}{\pi
m}(n+\frac{1}{2})\int_0^\infty \frac{x^2d
x}{(\omega^2-x^2)^2+\frac{\beta^2}{m^2}x^2}.\nonumber\\
&&
\end{eqnarray}
For sufficiently weak damping that is when $\beta $ is very small
, the integrands in (\ref{d17}) have singularity points $
x=\pm(\omega_1\pm\frac{i\beta}{2m})$ and by using  residual
calculus we find
\begin{equation}\label{d17.5}
lim_{t\rightarrow\infty}[ \langle\psi(0)| _\omega\langle n|: H_B
:|\psi(0)\rangle] =(n+\frac{1}{2})\omega
\end{equation}
Comparing (\ref{d16}) and (\ref{d17.5}), one can show that the
total energy of oscillator has been transmited to the reservoir
and according to (\ref{d16.5}), the kinetic energy of oscillator
tends to zero.\\
 If the state of system  in $ t=0 $ is $
\rho(0)=\rho_B^T\otimes|S\rangle_\omega $ where $
\rho_B^T=\frac{e^{\frac{-H_B}{KT}}}{Tr_B(e^{\frac{-H_B}{KT}})} $
is the Maxwell-Boltzman distribution and $ |S\rangle_\omega $ is
an arbitrary state of harmonic oscillator , then by using of $
Tr_B[b_{\vec{k}}^\dag(0)b_{\vec{k'}}(0)\rho_B^T]=\frac{\delta(\vec{k}-\vec{k'})}{e^{\frac{\omega_{\vec{k}}}{KT}}-1}
$ one can show the expectation value of kinetic energy of
oscillator in $ t\rightarrow\infty $ tends to
\begin{eqnarray}\label{d11.1}
&&lim_{t\rightarrow\infty} \langle
:\frac{1}{2}m\dot{q}^2(t)+\frac{1}{2}m\omega^2 q^2(t) : \rangle
=\frac{2\beta}{\pi m^2}\int_0^\infty
\frac{x}{[(\omega^2-x^2)^2+\frac{\beta^2}{m^2}x^2](e^{\frac{x}{KT}}-1)}dx\nonumber\\
&&+\frac{2\beta}{\pi m^2}\int_0^\infty
\frac{x^3}{[(\omega^2-x^2)^2+\frac{\beta^2}{m^2}x^2](e^{\frac{x}{KT}}-1)}dx
\end{eqnarray}
\subsection{Transition probabilities}
We can write the Hamiltonian (\ref{d1}) as
\begin{eqnarray}\label{d23}
&&H=H_0+H',\nonumber\\
&&H_0=(a^\dag a+\frac{1}{2})\omega+H_B, \nonumber\\
&&H'=-\frac{p}{m}R+\frac{R^2}{2m},
\end{eqnarray}
 where $ a $ and $ a^\dag $ are annihilation and creation operators of the Harmonic oscillator.
 In interaction picture we can write
\begin{eqnarray}\label{d24}
&& a_I(t)=e^{iH_0 t}a(0)e^{-iH_0 t}=ae^{-i\omega t},\nonumber\\
&& b_{\vec{k}I}(t)=e^{iH_0 t}b_{\vec{k}}(0)e^{-iH_0
t}=b_{\vec{k}}(0)e^{-i\omega_{\vec{k}} t},
\end{eqnarray}
the terms $ \frac{R}{m} p $ and $ \frac{R^2}{2m} $ are of the
first order and second order of damping respectively, therefore,
for a sufficiently weak damping, $ \frac{R^2}{2m}$ is small in
comparison with $ \frac{R}{m} p $. Furthermore  $ \frac{R^2}{2m}
$ has not any role in those transition probabilities where
initial and final states of harmonic oscillator are different,
hence we can neglect the term $\frac{R^2}{2m} $  in $ H' $.
Substituting $ a_I $ and $ b_{\vec{k}I} $ from (\ref{d24}) in $
-\frac{ R}{m} p $, one can obtain $ H_I' $ in interaction picture
as
\begin{eqnarray}\label{d25}
&&H_I'=-i\sqrt{\frac{\omega}{2m}}\int_{-\infty}^{+\infty}d^3 k
(f(\omega_{\vec{k}}) a^\dag b_{\vec{k}}(0)
e^{i(\omega-\omega_{\vec{k}})t}-f^*(\omega_{\vec{k}}) a
b_{\vec{k}}^\dag(0) e^{-i(\omega-\omega_{\vec{k}}
)t}\nonumber\\
&&-f(\omega_k) a b_{\vec{k}}(0)
e^{-i(\omega_{\vec{k}}+\omega)t}+f^*(\omega_k) a^\dag
b_{\vec{k}}(0) e^{i(\omega_{\vec{k}}+\omega) t}),
\end{eqnarray}
 the terms  containing just $
a b_{\vec{k}}(0) $ and $ a^\dag b_{\vec{k}}^\dag(0) $ violate the
conservation of energy in the first order perturbation, because $
a b_{\vec{k}}(0) $ destroys an excited state of harmonic
oscillator while at the same time destroying a reservoir
excitation state and $ a^\dag b_{\vec{k}}^\dag(0) $ creates an
excited state of harmonic oscillator, while creating an excited
reservoir state at the same time, therefore we neglect the terms
involving $ a b_{\vec{k}}(0) $ and $ a^\dag b_{\vec{k}}^\dag(0) $
because of energy conservation and write $ H'_I $ as
\begin{equation}\label{d26}
H_I'=-i\sqrt{\frac{\omega}{2m}}\int_{-\infty}^{+\infty}d^3 k
[f(\omega_{\vec{k}}) a^\dag b_{\vec{k}}(0)
e^{i(\omega-\omega_{\vec{k}})t}-f^*(\omega_{\vec{k}}) a
b_{\vec{k}}^\dag(0) e^{-i(\omega-\omega_{\vec{k}} )t}].
\end{equation}
Now the time evolution of density operator in interaction picture
is [14]
\begin{equation}\label{d27}
\rho_I(t)=U_I(t,t_0)\rho_I(t_0)U_I^\dag(t,t_0),
\end{equation}
where $ U_I $ is the time evolution operator, which in first
order perturbation is
\begin{eqnarray}\label{d28}
 &&U_I(t,t_0=0)=1-i \int_0^t d t_1 H'_I(t_1)=\nonumber\\
 && 1-\sqrt{\frac{\omega}{2m}}\int_{-\infty}^{+\infty}d^3 k
[f(\omega_{\vec{k}}) a^\dag b_{\vec{k}}(0)
e^{\frac{i(\omega-\omega_{\vec{k}})
t}{2}}\nonumber\\
&&-f^*(\omega_{\vec{k}}) a b_{\vec{k}}^\dag(0)
e^{\frac{-i(\omega-\omega_{\vec{k}})t}{2}}]
\frac{\sin\frac{(\omega-\omega_{\vec{k}})}{2}t}{\frac{(\omega-\omega_{\vec{k}})}{2}}.
\end{eqnarray}
Let $ \rho_I(0)=|n\rangle_{\omega\hspace{0.20 cm}\omega}\langle
n|\otimes|0\rangle_{B\hspace{0.20 cm}B} \langle
 0|$ where $ |0\rangle_B $ is the vacuum state of the reservoir and $ |n\rangle_\omega $, an
 excited state of the harmonic oscillator, then by substituting $ U_I(t,0) $ from
 (\ref{d28}) in (\ref{d27}) and taking trace  over reservoir parameters we obtain
 \begin{eqnarray}\label{d28.1}
 &&\rho_{sI}(t):=Tr_B(\rho_I(t))=|n\rangle_{\omega\hspace{0.20
cm}\omega}\langle n|\nonumber\\
&&+\frac{n \omega}{2m}|n-1\rangle_{\omega\hspace{0.20
cm}\omega}\langle n-1|\int_{-\infty}^{+\infty} d^3 p
|f(\omega_{\vec{p}})|^2
\frac{\sin^2\frac{(\omega_{\vec{p}}-\omega)}{2}t}
{(\frac{\omega_{\vec{p}}-\omega}{2})^2},\nonumber\\
&&
\end{eqnarray}
where we have used the formula $ Tr_B [ |1_{\vec{k}}\rangle_B
\hspace{00.20cm}_B\langle1_{\vec{k'}}|
]=\delta(\vec{k}-\vec{k'})$. For very large time, we can write
$\frac{\sin^2\frac{(\omega_{\vec{p}}-\omega)}{2}t}{(\frac{\omega_{\vec{p}}-\omega}{2})^2}=2\pi
t\delta(\omega_{\vec{p}}-\omega)$ which leads to the following
relation for density matrix
\begin{equation}\label{d28.2}
\rho_{sI}(t)=|n\rangle_{\omega\hspace{0.20 cm}\omega}\langle
n|+\frac{4\pi^2\omega^3 n
t|f(\omega)|^2}{m}|n-1\rangle_{\omega\hspace{0.20
cm}\omega}\langle n-1|,
\end{equation}
from density matrix we can calculate the probability of transition
$|n\rangle_\omega\rightarrow|n-1\rangle_\omega $ as
\begin{eqnarray}\label{d29}
&&\Gamma_{n\rightarrow n-1}=Tr [(|n-1\rangle_{\omega\hspace{0.20
cm}\omega}\langle
n-1|)\rho(t)]=\nonumber\\
&&Tr_s[(|n-1\rangle_{\omega\hspace{0.20 cm}\omega}\langle
n-1|)\rho_{sI}(t)]=\frac{4\pi^2 \omega^3 n t |f(\omega)|^2}{m},
\end{eqnarray}
where $ Tr_s $ denotes taking trace over harmonic oscillator
eigenstates. For the special choice (\ref{d10}), above transition
probability becomes
\begin{equation}\label{d30}
\Gamma_{n\rightarrow n-1}=\frac{n\beta t}{m}.
\end{equation}
 Now consider the case where the reservoir is an excited state in $ t=0$ for example $
\rho_I(0)=|n\rangle_{\omega\hspace{0.20 cm}\omega}\langle
n|\otimes|1_{\vec{p}_1},...1_{\vec{p}_j}\rangle_{B\hspace{0.20
cm}B} \langle
 1_{\vec{p}_1},...1_{\vec{p}_j}| $ where $ |1_{\vec{p}_1},...1_{\vec{p}_j}\rangle_B $ denotes a state of
 reservoir that contains $ j $ quanta with corresponding momenta $ \vec{p}_1,...\vec{p}_j $,
 then by making use of
 \begin{eqnarray}\label{d31}
 &&Tr_B[ b_{\vec{k}}^\dag |1_{\vec{p}_1},...1_{\vec{p}_j}\rangle_{B\hspace{0.20 cm}B} \langle
 1_{\vec{p}_1},...1_{\vec{p}_j}| b_{\vec{k'}}]=\delta(\vec{k}-\vec{k'}),\nonumber\\
 &&Tr_B[b_{\vec{k}} |1_{\vec{p}_1},...1_{\vec{p}_j}\rangle_{B\hspace{0.20 cm}B} \langle
 1_{\vec{p}_1},...1_{\vec{p}_j}|b_{\vec{k'}}^\dag ]=
 \sum_{l=1}^j \delta(\vec{k}-\vec{p}_l)\delta(\vec{k'}-\vec{p}_l),
 \end{eqnarray}
 and long time approximation, we find
 \begin{eqnarray}\label{d32}
 &&\rho_{sI}(t)=|n\rangle_{\omega\hspace{0.20 cm}\omega}\langle
n|+\frac{(n+1)\omega}{2m}|n+1\rangle_{\omega\hspace{0.20
cm}\omega}\langle n+1| \sum_{l=1}^j |f(\omega_{\vec{p}_l})|^2
\frac{\sin^2 \frac{(\omega_{\vec{p}_l}-\omega)}{2}t}{(
\frac{\omega_{\vec{p}_l}-\omega}{2})^2}\nonumber\\
&&+\frac{n\omega}{2m}|n-1\rangle_{\omega\hspace{0.20
cm}\omega}\langle n-1|\int_{-\infty}^{+\infty} d^3 k
|f(\omega_{\vec{k}})|^2 \frac{\sin^2
\frac{(\omega_{\vec{k}}-\omega)}{2}t}{(
\frac{\omega_{\vec{k}}-\omega}{2})^2},
\end{eqnarray}
which gives the transition probability for $ |n\rangle_\omega
\rightarrow|n-1\rangle_\omega $ and $|n\rangle_\omega\rightarrow
|n+1\rangle_\omega $, respectively as follows
\begin{eqnarray}\label{d32}
&&\Gamma_{n\rightarrow n-1}=Tr_s[|n-1\rangle_{\omega\hspace{0.20
cm}\omega}\langle n-1| \rho_{sI}(t)]=\frac{4\pi^2
\omega^3 n t}{m}|f(\omega)|^2,\nonumber\\
&&\Gamma_{n\rightarrow n+1}=Tr_s[|n+1\rangle_{\omega\hspace{0.20
cm}\omega}\langle n+1| \rho_{sI}(t)]=\frac{(n+1)\pi
t\omega}{m}|f(\omega)|^2 \sum_{l=1}^j \delta(
\omega_{\vec{p}_l}-\omega).\nonumber\\
&&
\end{eqnarray}
Specially for the choice (\ref{d10}), we have
\begin{eqnarray}\label{d33}
&&\Gamma_{n\rightarrow n-1}=\frac{n\beta
t}{m},\nonumber\\
&&\Gamma_{n\rightarrow n+1}=\frac{\beta(n+1) t}{4\pi m \omega^2}
\sum_{l=1}^j \delta( \omega_{\vec{p}_l}-\omega).
\end{eqnarray}
Another important case is when the reservoir has a
Maxwell-Boltzman distribution so let
$\rho_I(0)=|n\rangle_{\omega\hspace{0.20 cm}\omega}\langle n|
\otimes \rho_B^T $ where\\  $ \rho_B^T=\frac{e^{\frac{-H_B}{K
T}}}{TR_B(e^{\frac{-H_B}{KT}})} $, then by making use of following
relations
\begin{eqnarray}\label{d34}
&&Tr_B[ b_{\vec{k}}\rho_B^T b_{\vec{k'}}]=Tr_B[ b_{\vec{k}}^\dag
\rho_B^T
b_{\vec{k'}}^\dag]=0,\nonumber\\
&& Tr_b[b_{\vec{k}}\rho_B^T
b_{\vec{k'}}^\dag]=\frac{\delta(\vec{k}-\vec{k'})}
{e^{\frac{\omega_{\vec{k}}}{K T}}-1},\nonumber\\
&&Tr_B[ b_{\vec{k}}^\dag \rho_B^T
b_{\vec{k'}}]=\frac{\delta(\vec{k}-\vec{k'})e^{\frac{\omega_{\vec{k}}}{K
T}}}{e^{\frac{\omega_{\vec{k}}}{K T}}-1},
\end{eqnarray}
we can obtain the density operator $ \rho_{sI}(t) $ in interaction
picture as
 \begin{eqnarray}\label{d35}
 &&\rho_{sI}(t):=Tr_B [\rho_I(t)]=|n\rangle_{\omega\hspace{0.20 cm}\omega}\langle
n|\nonumber\\
&&+\frac{(n+1)\omega}{2m}|n+1\rangle_{\omega\hspace{0.20
cm}\omega}\langle n+1| \int_{-\infty}^{+\infty} d^3 k
\frac{|f(\omega_{\vec{k}})|^2}{e^{\frac{\omega_{\vec{k}}}{K
T}}-1} \frac{\sin^2 \frac{(\omega_{\vec{k}}-\omega)}{2}t}{
(\frac{\omega_{\vec{k}}-\omega}{2})^2}\nonumber\\
&&+\frac{n\omega}{2m}|n-1\rangle_{\omega\hspace{0.20
cm}\omega}\langle n-1|\int_{-\infty}^{+\infty} d^3 k
\frac{|f(\omega_{\vec{k}})|^2e^{\frac{\omega_{\vec{k}}}{K
T}}}{e^{\frac{\omega_{\vec{k}}}{K T}}-1} \frac{\sin^2
\frac{(\omega_{\vec{k}}-\omega)}{2}t}{(
\frac{\omega_{\vec{k}}-\omega}{2})^2},
\end{eqnarray}
which accordingly gives the following transition probabilities in
verey long time as
\begin{eqnarray}\label{d35}
&&\Gamma_{n\rightarrow n-1}=Tr_s[|n-1\rangle_{\omega\hspace{0.20
cm}\omega}\langle n-1| \rho_{sI}(t)]=\frac{4\pi^2\omega^3 n
t}{m}\frac{|f(\omega)|^2 e^{\frac{\omega}{K
T}}}{e^{\frac{\omega}{K T}}-1},\nonumber\\
&&\Gamma_{n\rightarrow n+1}=Tr_s[|n+1\rangle_{\omega\hspace{0.20
cm}\omega}\langle n+1| \rho_{sI}(t)]=\frac{4\pi^2\omega^3 (n+1)
t}{m}\frac{|f(\omega)|^2}{e^{\frac{\omega}{K T}}-1},\nonumber\\
&&
\end{eqnarray}
substituting (\ref{d10}) in these recent relations we find
\begin{eqnarray}\label{d36}
&&\Gamma_{n\rightarrow n-1}=\frac{n\beta te^{\frac{\omega}{K
T}}}{m(e^{\frac{\omega}{K T}}-1)},\nonumber\\
&&\Gamma_{n\rightarrow n+1}=\frac{(n+1)\beta
t}{m(e^{\frac{\omega}{K T}}-1)}.
\end{eqnarray}
 So in very low temperatures the energy flows from oscillator to the reservoir by the rate
  $ \Gamma_{n\rightarrow n-1}\mapsto\frac{n\beta}{m}$ and no energy flows from
  reservoir to oscillator.
  \section{Quantum field of reservoir}
Let us define the operators $ Y(x,t) $ and $ \Pi_Y(x,t) $ as
follows
\begin{eqnarray}\label{d18}
&&Y(\vec{x},t)=\int_{-\infty}^{+\infty} \frac{d^3
k}{\sqrt{2(2\pi)^3\omega_{\vec{k}}}}(
b_{\vec{k}}(t)e^{i\vec{k}.\vec{x}}+b_{\vec{k}}^\dag(t)e^{-i\vec{k}.\vec{x}}),\nonumber\\
&&\Pi_Y(\vec{x},t)=i\int_{-\infty}^{+\infty} d^3 k
\sqrt{\frac{\omega_{\vec{k}}}{2(2\pi)^3}}( b_{\vec{k}}^\dag
(t)e^{-i\vec{k}.\vec{x}}-b_{\vec{k}}(t)e^{i\vec{k}.\vec{x}}),
\end{eqnarray}
then using commutation relations (\ref{d4}), one can show that $
Y(\vec{x},t) $ and $ \Pi_Y (\vec{x},t)$, satisfy the equal time
commutation relations
\begin{equation}\label{d19}
[ Y(\vec{x},t),\Pi_Y(\vec{x'},t)]=i\delta(\vec{x}-\vec{x'}),
\end{equation}
furthermore by substituting $ b_k(t)$ from ( \ref{d8}) in
(\ref{d18}), we obtains
\begin{eqnarray}\label{d20}
&&\frac{\partial\Pi_Y(\vec{x},t)}{\partial
t}=\nabla^2Y+2\dot{q}(t) P(\vec{x}),\hspace{1.00 cm
}P(\vec{x})=Re{ \int_{-\infty}^{+\infty}
d^3 k\sqrt{\frac{\omega_{\vec{k}}}{2(2\pi)^3}}f(\omega_{\vec{k}})e^{-i\vec{k}.\vec{x}}},\nonumber\\
&&\Pi_Y(\vec{x},t)=\frac{\partial Y}{\partial t}-2\dot{q}(t)
Q(\vec{x}),\hspace{1.00 cm} Q(\vec{x})= I
m{\int_{-\infty}^{+\infty} d^3
k\frac{f(\omega_{\vec{k}})}{\sqrt{2(2\pi)^3\omega_{\vec{k}}}}e^{-i\vec{k}.\vec{x}}},\nonumber\\
&&
\end{eqnarray}
so $ Y(\vec{x},t)$ satisfies the following source included
Klein-Gordon equation
\begin{equation}\label{d21}
\frac{\partial^2Y}{\partial
t^2}-\nabla^2Y=2\ddot{q}(t)Q(\vec{x})+2\dot{q}(t)P(\vec{x}),
\end{equation}
with the corresponding Lagrangian density
\begin{equation}\label{d22}
\pounds=\frac{1}{2}(\frac{\partial Y}{\partial
t})^2-\frac{1}{2}\vec{\nabla Y}.\vec{\nabla
Y}-2\dot{q}Q(\vec{x})\frac{\partial Y}{\partial
t}+2\dot{q}P(\vec{x})Y.
\end{equation}
Therefore the reservoir is a massless Klein-Gordon field with
source $ 2\ddot{q}Q(\vec{x})+2\dot{q}P(\vec{x})$. The Hamiltonian
density for (\ref{d21}) is as follows
\begin{equation}\label{d22.5}
\aleph=\frac{(\Pi_Y+2\dot{q}Q)^2}{2}+\frac{1}{2}|\vec{\nabla
Y}|^2-2\dot{q}Y P,
\end{equation}
and equations (\ref{d20}) are Heisenberg equations for $ Y $ and
$ \Pi_Y$. If we obtain $ b_{\vec{k}} $ and $ b_{\vec{k}}^\dag $
from (\ref{d18}) in terms of $ Y $ and $ \Pi_Y $  and substitute
them in $ H_B $ defined in (\ref{d3})we obtain
\begin{equation}\label{d22.75}
H_B=\int_{-\infty}^{+\infty}d^3k \omega_{\vec{k}}b_{\vec{k}}^\dag
b_{\vec{k}}=\frac{\Pi_Y^2}{2}+\frac{1}{2}|\vec{\nabla Y }|^2.
\end{equation}
\section{Concluding remarks}
 By generalizing Caldeira-Legget model to an environment with
 continuous degrees of freedom, for example a Klein-Gordon field,
 a new minimal coupling method introduced which can be extended and applied to a
 large class of dissipative systems consistently.
 Such method applied to a quantum damped harmonic oscillator as a prototype of these
 models, with a dissipation term proportional to velocity. Some
 transition probabilities explaining the way energy flows between
 subsystems obtained. Choosing different coupling functions in
 (\ref{d5}) we could investigate another classes of dissipative systems.

\end{document}